
\input phyzzx
\FRONTPAGE
\line{\hfill BROWN-HET-960}
\line{\hfill July 1994}
\bigskip
\titlestyle{{\bf FUNCTIONAL APPROACH TO STOCHASTIC INFLATION} \foot{Work
supported in part by the Department of Energy under contract
DE-FG02-91ER40688 - Task A}\break}
\author{Yuri V. Shtanov}
\bigskip
\address{Bogolyubov Institute for Theoretical Physics, Kiev, 252143, Ukraine
\footnote{\dagger}{\rm Permanent address}}
\address{and}
\address{Department of Physics, Brown University, Providence, RI 02912, USA}
\abstract

 We propose functional approach to the stochastic inflationary universe
dynamics. It is based on path integral representation of the solution
to the differential equation for the scalar field probability distribution.
In the saddle-point approximation scalar field probability distributions
of various type
are derived and the statistics of the inflationary-history-dependent
functionals is developed.
\endpage

{\bf \chapter{Introduction}}

There  is  an  undecreasing   interest  in  the stochastic
dynamics of  the inflationary universe (see \REF\llm{A.Linde,
D.Linde and A.Mezhlumian, {\it Phys. Rev.} {\bf D49} (1994) 1783.} [\llm]
for the most recent developments). It is now well known \Ref\l{A.D.Linde
{\it Particle
Physics and Inflationary Cosmology} (Harwood, Chur, Switzerland, 1990)}
that during inflation the inflaton scalar
field is subject to quantum fluctuations and
for this reason its behaviour under certain conditions resembles stochastic
Brownian motion rather than regular
evolution. As it  has  been  discussed about ten  years  ago
\Ref\s{A.A.Starobinsky,  in   {\it Fundamental
interactions},   (MGPI   Press,
Moscow, 1983) p.55; \nextline  A.A.Starobinsky, in  {\it Current Topics
in   Field
Theory, Quantum Gravity and Strings}, {\it Lecture Notes in Physics},  eds.
H.J. de Vega and  N.Sanchez, (Springer-Verlag,  Heidelberg, 1986) Vol.
{\bf 246}, p.107.}
such a stochastic behaviour can be described by a probability distribution
$P(\phi, \tau)$ of the inflaton
field values $\phi$ in a Hubble-size domain at the inflationary stage.
Here $\tau$ denotes an arbitrary timelike evolution parameter which will be
specified in a moment. More precisely, the function $P(\phi, \tau)$ is the
probability density for a hypothetical observer at the inflationary stage
to find in its Hubble-size vicinity at the moment of ``time'' $\tau$
the average value of the scalar field close to $\phi$.
This probability distribution evolves with ``time'' $\tau$ and obeys the
following differential equation [\s]
$$
{\partial P(\phi, \tau) \over \partial \tau} \ = \ {\partial \over \partial
\phi} \left( A(\phi) P(\phi, \tau) \, + \, {1 \over 2} B^{1 - \alpha} (\phi)
{\partial \over \partial \phi}
B^\alpha (\phi) P(\phi, \tau) \right) \, , \eqno \eq
$$
where the constant parameter $\alpha$ reflects some ambiguity in the
order of the differential operations on the right-hand side (see e.g. [\llm]).
The  functions $A(\phi)$ and $B(\phi)$  depend on the specific choice of the
parameter $\tau$. In this paper we shall consider two such choices. Namely,
if $\tau$  is  the  cosmological  time, $\tau = t$, then
$$
A(\phi) \ = \ {1 \over 3H(\phi)}\, {d V(\phi) \over d \phi}
\, , \hskip2cm
B(\phi) \ = \ {H^3(\phi) \over 4 \pi^2} \, ,  \eqno \eq
$$
If $\tau$ is the expansion power, i.e.  $\tau = \log a$, where $a$ is the
scale factor of the universe, then
$$
A(\phi) \ = \ {1 \over 3H^2(\phi)}\, {d V(\phi) \over d \phi}
\, , \hskip1.8cm
B(\phi) \ = \ {H^2(\phi) \over 4 \pi^2} \, , \eqno \eq
$$
In  these  expressions $H(\phi)$ is the  Hubble parameter which on the
inflatonary stage is given to a good precision by
$$
H^2(\phi) = {8\pi \over 3M_{\rm P}^2}\, V(\phi)\, , \eqno \eq
$$
$V(\phi)$ is the scalar field potential, and $M_{\rm P}$
is  the  Planck  mass. In all the cases for simplicity we call the
parameter $\tau$ ``time.''

The applicability limits of Eq. (1.1) were under consideration in [\s] (see
also \REF\glm{A.S.Goncharov, A.D.Linde and V.F.Mukhanov,
{\it Int. J.  Mod. Phys.}  {\bf A2} (1987) 561.} [\glm, \llm]).
We shall discuss this point in the following chapter. Now it will be
sufficient to note that for any  reasonable inflaton
potential there is a wide range of the scalar  field  values  for  which
this equation is valid.

Investigation of Eq. (1.1) and search of  its  approximate  solutions
was  the  subject  of  many  works (see [\llm] and references therein).
The  aim   of this  paper  is   to propose an  approach  based    on
functional  calculus. As we will see, this  approach  is   fairly
transparent   and  sometimes
turns  out  to  be very  convenient.    Moreover,    it    represents
a  useful tool when studying statistics of the
inflationary-history-dependent quantities. Such
an  observable value has been proposed  in  the
work \REF\cs{G.V.Chibisov and Yu.V.Shtanov, {\it Int. J. Mod. Phys.} {\bf A5}
(1990) 2625.} [\cs], in which also the approach which will
be developed in this paper was partially used.

Our  starting point  is  to write down  the formal solution  to (1.1)
in  terms  of  path  integral
$$
Z(\phi_f, \phi_i, \tau) \ = \ {\cal N} \int \exp \bigl( - I[\phi(\tau)]
\bigr)\,
[d\phi] \, . \eqno \eq
$$
The function $Z(\phi_f,\phi_i,\tau)$ represents
the probability distribution of the transition
from  the initial scalar field  value  $\phi_i$  to the final value $\phi_f$
in  a  time  $\tau$,
${\cal N}$ here and below denotes an unspecified normalization constant,
$$
I[\phi(\tau)] \ = \ {1 \over 2} \int {\left(\dot \phi + A(\phi) \right)^2
\over B(\phi)}\, d\tau \, , \eqno \eq
$$
is the ``action''. The integral in (1.5) is taken over paths that
lead  from $\phi_i$ to $\phi_f$  in  a  time $\tau$. Overdot denotes the
derivative with respect to $\tau$. The integration measure $[d\phi]$ can be
formally
expressed as
$$
[d\phi] \ = \ \prod_\tau {d \phi(\tau) \over \sqrt{B(\tau)}} \, . \eqno \eq
$$
We shall not be rigorous and shall not specify the meaning of the path integral
(1.5) and of the measure (1.7).
We only note that it depends on the specification of the operator ordering
in (1.1).

The  representation  of the  solution $Z(\phi_f, \phi_i, \tau)$  in
          the  form  (1.5) makes it possible to interprete the expression
          $\exp (- I[\phi(\tau)])$  as the  probability  density in the space
of paths. Such an interperetation becomes more  evident  if
          one  considers  the  problem  of  calculating  mean values of
          functionals $F[\phi(\tau)]$ under  the  condition that the scalar
          field  evolution   started  at  $\phi_i$, ended  at $\phi_f$
and lasted for the  time  $\tau$. The mean value of the
 functional  $F[\phi(\tau)]$  will  be given  by  the  path integral similar
to (1.5)
$$
<F[\phi(\tau)]> \ = \ {\cal N} \int F[\phi(\tau)] \exp \left(- I[\phi(\tau)]
\right)\, [d \phi] \, .  \eqno \eq
$$
Path  integration  method  can  be  used  in  deriving   the
statistical properties of such functionals. In Chapter 5 we will  see  how
it works.

Of course in most cases it is impossible to calculate the path integral
   (1.5)  or  (1.8)  explicitly  as well as  it  is  impossible  to
  explicitly  solve  the  differential  equation  (1.1). Therefore we
  must look  for approximate solutions. In  this  paper   we
shall consider the case when the saddle-point approximation  is
good so that instead of (1.5) we can write approximately
$$
Z(\phi_f, \phi_i, \tau) \ \approx \ {\cal N} \exp \left(- I(\phi_f, \phi_i,
\tau) \right) \, , \eqno \eq
$$
where $I(\phi_f, \phi_i, \tau)$ is  the action on the extremal path.
The   applicability   limits  of  such  an   approximation   will    be
studied in the following chapter. We  will  see  that  these  limits  are
 not at all restrictive, and, in fact, reduce to
$$
V(\phi) \ \ll \ M_{\rm P}^4 \,    \eqno \eq
$$
for all the values of $\phi$ along the extremal path.
The condition  (1.10)  just  meets  the  condition  of  validity
of  Eq. (1.1) itself (see [\s, \glm, \llm]).
Thus we shall conclude that  to
the extent to which Eq. (1.1) is valid its approximate  solution  (1.9)  is
also valid. This conclusion will be justified also in Chapter 4 in which we
consider a special case of quartic potential $V(\phi)$ which allows for
analytic solution. This circumstance makes the saddle-point approximation
representative enough.

In Chapter 3 we  shall
consider the scalar field probability  distributions  of  various
types. We will  notice  dramatic
dependence  of  the  results  on  the specific choice of the parameter $\tau$
and on the type of  the
distribution  we consider. In Chapter 4, as we already mentioned, we
consider a special case of quartic potential $V(\phi)$ with
parameter $\tau$ being cosmological time, which can be solved exactly.
In Chapter 5 we shall develop a  general  method
of  investigation  of  the   path   functionals   statistics   in   the
saddle-point   approximation.  In  the final
Chapter 6  we shall briefly discuss possible applications of the tools
developed.

{\bf \chapter{Validity limits of the saddle-point  approximation}}

First of all it will be convenient to rewrite the action (1.6)  as follows
$$
I[\phi(\tau)] \ = \ S(\phi_f, \phi_i)  +  I_0 [\phi(\tau)] \, , \eqno \eq
$$
where \footnote{1)}{This and similar expressions resemble the wave function
of the universe in the de Sitter minisuperspace model. The authors of [\glm]
and [\llm] regard this fact rather seriously and in its view try
to relate
minisuperspace quantum cosmology to the stochastic approach to inflation.
To us it seems that this coincidence is but occasional.}
$$
S(\phi_f, \phi_i) \ = \ \int \limits_{\phi_i}^{\phi_f} {A(\phi) \over
B(\phi)}\, d \phi \ = \ {3M_{\rm P}^4 \over 16}\, \left( {1 \over
V(\phi_i)} - {1 \over V(\phi_f)} \right)\, , \eqno \eq
$$
and
$$
I_0[\phi(\tau)] \ = \ {1 \over 2} \int {\dot \phi^2 + A^2(\phi) \over B(\phi)}
\, d \tau \, . \eqno \eq
$$

 From the action (2.3) one finds the generalized momentum
$$
p \ = \ {\dot \phi \over B(\phi)} \, . \eqno \eq
$$
Our experience with quantum mechanics tells us that the saddle-point
(semiclassical)
approximation is applicable when the functions $A(\phi)$
and $B(\phi)$ do not change significantly at one ``wavelength'' distance
$l \sim |p|^{- 1}$ determined by the momentum $p$ of the extremal trajectory.
As both $A(\phi)$ and $B(\phi)$ depend  only on the potential $V(\phi)$
this  condition reduces  to
$$
l\, {d \log V(\phi) \over d \phi} \ \ll \ 1 \, . \eqno \eq
$$
Assuming semiclassical approximation to be valid we note that in this case the
peak of the distribution
(1.5)
at the moment $\tau$ will lie close to the value $\phi(\tau)$ of the extremal
trajectory which obeys $\dot \phi = - A(\phi)$. This follows from the
observation that the action (1.6) achieves its minimum on this trajectory.
Hence the condition (2.5) is to be checked on this trajectory.
Using the expressions (1.2) and (1.3) for $A(\phi)$ and $B(\phi)$ we obtain
$$
H^4(\phi) \ \ll \ V(\phi) \, , \eqno \eq
$$
which is equivalent to (1.10).

The  expression (1.10) is very notable: it  is  the  condition  under
which the universe can be treated as being classical. It is of a pleasant
surprise that in determining the validity limits of the saddle-point
(semiclassical)
approximation to the scalar field dynamics  we  ultimately  came  to  this
 condition. Futhermore,  the  condition  (1.10)  is the same as the  validity
condition  of the equation (1.1) itself (see [\s, \glm, \llm]). All this
enables us  to  think that the  saddle-point approximation  considered in  this
 paper  is
representative enough and that the methods developed in its context  can
be applied to a wide class of problems.

{\bf \chapter{Scalar field probability distributions}}

\section{Direct (in time) transition probability}

The transition probability distribution  is  given
by Eq. (1.5),  and, in  the  saddle-point  approximation, by
Eq. (1.9). From Eq. (1.6) it is clear that the peak of the distribution (1.9)
is at $\phi_f = \phi_d (\tau)$, where $\phi_d (\tau)$  is the direct (in time)
extremal
path with the lowest possible action, that is, the solution to the equation
$$
{d \phi \over d \tau} \ = \ - \ A(\phi) \, , \eqno \eq
$$
with the initial condition $\phi(0) = \phi_i$.

We  can  develop the function $I(\phi_f,\phi_i,\tau)$ in  powers  of
$\phi_f - \phi_d (\tau)$ as follows
$$
I(\phi_f, \phi_i, \tau) \ = \ {\left( \phi_f - \phi_d(\tau) \right)^2 \over
2!\, \Delta_d(\tau)} + {\left( \phi_f - \phi_d(\tau) \right)^3 \over
3!\, \Gamma_d(\tau)} +  \ldots \, , \eqno \eq
$$
and restrict  ourselves  to terms  quadratic  in $\phi_f - \phi_d (\tau)$.
This assumes the distribution to be close to Gaussian. The expression for
the value $\Delta_d(\tau)$ is easily obtained by making use of
the Hamilton-Jacobi equation for the function $I(\phi_f, \phi_i, \tau)$.
The result is
$$
\Delta_d \ = \ A^2(\phi_d) \Big| \int \limits_{\phi_i}^{\phi_d} {B(\phi)
\over A^3(\phi)}\, d\phi \Big| \, . \eqno \eq
$$
In the work [\glm] the same expression for the variance $\Delta_d$ of the
scalar
field
distribution in Gaussian approximation was obtained by a different method.

As an example, consider the case of the power law potential
$$
V(\phi) \ =  \ {\lambda \phi^n \over n M_{\rm P}^{n - 4}} \hskip2cm  {\rm (}n\
{\rm is\ even)} \, . \eqno \eq
$$
If $\tau$  is  the cosmological time then taking into account (1.2) we will
have
$$
\Delta_d \ = \ {4 \over 3} {\big| \phi_i^4 - \phi_d^4 \big| \over n^2
M_{\rm P}^n}\, \lambda \phi_d^{n - 2} \, . \eqno \eq
$$
If $\tau$ is the expansion power, $\tau = \log a$, we obtain using (1.3)
$$
\Delta_d \ = \ {16 \over 3} {\big| \phi_i^{n + 4} - \phi_d^{n + 4} \big|
\over n^2 (n + 4) M_{\rm P}^n}\, \lambda \phi_d^{- 2} \, . \eqno \eq
$$

The dramatic difference between the variances (3.5)  and
(3.6) can be seen  in their  behaviour at sufficiently small $\phi_d$. For
$ n
> 2$ if $\tau$ is the cosmological time then the variance, which  is  given
by (3.5), is decreasing with time when $\phi_d$ is sufficiently small.
The   fluctuations    thereby   always
remain    sharply  peaked  around  their  maximum.  Yet  if  $\tau$  is  the
expansion power the variance is increasing rather then  decreasing.
Surprisingly enough, the approaches to  the   scalar  field  statistics
that seem  to  be just slightly different bring so different results.

Natural question is: when is it correct to treat  the  fluctuations  as
Gaussian. The answer is that
it is possible to do so if the condition  (1.10)  is valid  (see  the
work \REF\h{H.M.Hodges,  {\it Phys. Rev.} {\bf D39} (1989) 3568.} [\h] and
Chapter 4 of the present paper
where this is shown  for  the exactly solvable case  of $\lambda \phi^4$
potential).  We
remember (this has been shown in Chapter 2) that under this condition also
the  saddle-point   approximation  is  valid.
The  condition  (1.10) thus turns out  to be sufficient for all
the assumptions made in this paper.
\smallskip

\section{Reverse (in time) transition probability}

    Besides the direct (in time) transition probability we can consider the
probability  that  the  scalar field initially had  the  value $\phi_i$ if
after  the  time  $\tau$  it  has  the value $\phi_f$. The equation for this
probability is obtained from (1.1) by replacing the differential operator
in the right-hand side by its conjugate [\llm]
$$
{\partial P(\phi, \tau) \over \partial \tau} \ = \ - \ A(\phi) {\partial
P(\phi, \tau)
\over \partial \phi} \ + \ {1 \over 2} B^\alpha (\phi) {\partial \over \partial
\phi} B^{1 - \alpha} (\phi) {\partial P(\phi, \tau) \over \partial \phi}
\, . \eqno \eq
$$
It can be shown that the solution to this equation which describes the
reverse (in time) transition from $\phi_i$ to $\phi_f$ is given by the same
function
$Z(\phi_f, \phi_i, \tau)$ represented by the
path integral (1.5). In  the saddle-point  approximation  this   probability
is  given by the equation (1.9).
The peak of the distribution (1.9), regarded as function of $\phi_i$, is at
$\phi_i = \phi_r(\tau)$,  where $\phi_r(\tau)$ is the reverse
extremal trajectory with minimal possible action, or solution to the equation
$$
 {d \phi \over d \tau} \ = \  A(\phi) \,  ,  \eqno \eq
$$
with the initial condition $\phi(0) = \phi_f$.

The  extremal  action in  (1.9) can be developed in powers of $\phi_i - \phi_r
(\tau)$ (similar to its development (3.2) in powers of $\phi_f - \phi_d$)
$$
I(\phi_f, \phi_i, \tau) \ = \ {\left( \phi_i - \phi_r(\tau) \right)^2 \over
2!\, \Delta_r(\tau)} + {\left( \phi_i - \phi_r(\tau) \right)^3 \over
3!\, \Gamma_r(\tau)} +  \ldots \, , \eqno \eq
$$
and the expression for $\Delta_r(\tau)$ is easily found to be
$$
\Delta_r \ = \ A^2(\phi_r) \Big| \int \limits_{\phi_f}^{\phi_r} {B(\phi)
\over A^3(\phi)}\, d\phi \Big| \, . \eqno \eq
$$
For the scalar field potential (3.4) in the case when $\tau$ is  the
cosmological  time we will have
$$
\Delta_r \ = \ {4 \over 3} {\big| \phi_r^4 - \phi_f^4 \big| \over n^2
M_{\rm P}^n}\, \lambda \phi_r^{n - 2} \, . \eqno \eq
$$
If $\tau$ is the expansion power, $\tau = \log a$, we obtain
$$
\Delta_r \ = \ {16 \over 3} {\big| \phi_r^{n + 4} - \phi_f^{n + 4} \big|
\over n^2 (n + 4) M_{\rm P}^n}\, \lambda \phi_r^{- 2} \, . \eqno \eq
$$
We see  that  in  both  cases  as $\phi_r \gg \phi_f$  the   behaviour
of  the variance  is
$$
\Delta_r \ \sim \ {V(\phi_r) \over M_{\rm P}^4}\, \phi_r^2 \, . \eqno \eq
$$
As $\phi_r^2$ grows  with time, the variance $\Delta_r$ is also  growing,
but  as long as $V(\phi_r) \ll M_{\rm P}^4$ the distribution  is  sharply
peaked  around  its maximum.
\smallskip

\section{Conditional probability distribution}

 In  most  of  the  works on the topic under consideration
 only simple direct or reverse transition probability distributions
were considered. However it
 seems even more interesting to inquire about the  conditional
 distributions.
Specifically, the distribution $Z(\phi_f, \phi_i, \tau_\ast; \phi, \tau)$
for the values of the scalar field $\phi$ at the moment of
time $\tau < \tau_\ast$  under the condition that the scalar field passes from
$\phi_i$ to $\phi_f$  in a time $\tau_\ast$  is given by
$$
Z(\phi_f, \phi_i, \tau_\ast; \phi, \tau) \ = \ {Z(\phi_f, \phi, \tau_\ast -
\tau) \cdot Z(\phi, \phi_i, \tau) \over Z(\phi_f, \phi_i, \tau_\ast)}
\, . \eqno \eq
$$
In Gaussian approximation this expression will look as follows
$$
Z(\phi_f, \phi_i, \tau_\ast; \phi, \tau) \ \approx \ {\cal N} \exp \left( -
{\left(\phi - \phi_m \right)^2 \over 2 \Delta_m} \right) \, , \eqno \eq
$$
where $\phi_m$ is the value of the scalar field $\phi$ at the peak of the
distribution, and $\Delta_m$ is the variance analogous to $\Delta_d$ and
$\Delta_r$ of the previous sections. If the value of $\tau_\ast$ is
sufficiently close to the amount of time it takes to proceed from $\phi_i$
to $\phi_f$ along the trajectory defined by (3.1), then the peaks of both
multipliers in the numerator of (3.14) at the moment $\tau < \tau_\ast$ will
be around the same value of $\phi \approx \phi_1(\tau) \approx \phi_2(\tau)$.
The values $\phi_1 (\tau)$ and $\phi_2 (\tau)$ are defined as follows:
$\phi_1 (\tau)$ is  the  starting  point of the extremal path determined by
(3.1) which ends  at
$\phi_f$ and which lasts for the time  $\tau_\ast - \tau$, and $\phi_2 (\tau)$
is the end point of the extremal path determined by (3.8) which starts at
$\phi_i$ and which lasts for the time $\tau$.
Then writing  for  both   multipliers  in  the   numerator  of
  Eq. (3.14)  the  expressions  similar to  (1.9) and
  developing the extremal actions in the way similar to  (3.2)
  and  (3.9)  we can obtain the following approximate expression
for (3.14)
$$
Z(\phi_f, \phi_i, \tau_\ast; \phi, \tau) \ \approx \ {\cal N} \exp \left( -
{\left(\phi - \phi_1 \right)^2 \over 2 \Delta_1} - {\left( \phi - \phi_2
\right)^2 \over 2 \Delta_2} \right) \, . \eqno \eq
$$
  The variances $\Delta_1$ and $\Delta_2$ are
given by the expressions similar to (3.3) and (3.10). From (3.16) one can see
that the values $\phi_m$ and $\Delta_m$ in (3.15) are given by
$$
\phi_m \ = \ {\phi_1 \Delta_2 + \phi_2 \Delta_1 \over \Delta_1 + \Delta_2}
\, ,\eqno \eq
$$
$$
\Delta_m \ = \ {\Delta_1 \Delta_2 \over \Delta_1 + \Delta_2} \, . \eqno \eq
$$

We can see that  the  peak $\phi_m$ of the  distribution (3.14) is
  between $\phi_1$ and $\phi_2$, and the variance $\Delta_m$ is smaller than
in the case of the unconditioned distribution (1.9). We must note, however,
that if the value of $\tau_\ast$ strongly deviates from the value of time it
takes to proceed from $\phi_i$ to $\phi_f$ along the trajectory defined by
(3.1), then the approximation (3.16) will be not good. In this case higher
order terms in the developments (3.2) and (3.9) of the extremal actions will
contribute significantly to the values of $\phi_m$ and $\Delta_m$.
In Chapter 5 the variance of
the conditional probability distribution will be calculated more precisely
using path functional statistics.

{\bf \chapter{Exact analytic solution}}

In the paper [\h] it was demonstrated that the stochastic equations can be
solved analytically for the theory with quartic potential
$$
V(\phi) \ = \ {\lambda \over 4}\, \phi^4 \, , \eqno \eq
$$
and in the case when $\tau$ is proportional to the cosmological time $t$.
Derivation in [\h] was based on the Langevin stochastic equation from which
the Fokker-Planck equation (1.1) is usually obtained (see the Appendix).
In this chapter we show that the case under consideration in our
formalism is described by a solvable path integral \footnote{1)}{Such a
possibility was communicated to us by V.Mukhanov.} (see also
\REF\m{V.F.Mukhanov,
Preprint ETH (1994), to appear.} Ref. \m). If we return to the
expressions (1.5)-(1.7), make a transformation from $\phi$ to a new variable
$$
x \ = \ \sqrt{3 \over 2 \lambda}\, \left({M_{\rm P} \over \phi} \right)^2
\, , \eqno \eq
$$
and (for convenience) rescale the time as follows
$$
\tau \ = \ \sqrt{2 \lambda \over 3 \pi}\, M_{\rm P}t \, , \eqno \eq
$$
then the expression (1.6) for the action $I$ will acquire a simple form
 $$
I[x(\tau)] \ = \ {1 \over 2} \int \left( \dot x - x \right)^2\, d \tau
\, , \eqno \eq
$$
and the expression similar to (1.5) will become
$$
Z(x_f, x_i, \tau) \ = \ {\cal N}
\int \exp \left( - I[x(\tau)] \right)\, [dx] \, , \eqno \eq
$$
with the standard measure
$$
[dx] \ = \ \prod_\tau dx(\tau) \, . \eqno \eq
$$

Path integral (4.5) with the quadratic action (4.4) can be calculated exactly
and one obtains the distribution for $x_f$ with fixed initial $x_i$
(direct transition probability) in the form
$$
Z(x_f, x_i, \tau) \ = \ {\cal N} \exp \left( - {\left(x_f - x_d (\tau)
\right)^2 \over \left( \left( x_d (\tau) / x_i \right)^2 - 1 \right) } \right)
\, .  \eqno \eq
$$
The same expression can be rewritten to yield the distribution for $x_i$ with
fixed final $x_f$ (reverse transition probability) as follows
$$
Z(x_f, x_i, \tau) \ = \ {\cal N} \exp \left( - {\left(x_i - x_r (\tau)
\right)^2 \over \left(1 - \left( x_r (\tau) / x_f \right)^2 \right) } \right)
\, .  \eqno \eq
$$
The definition of the functions $x_d(\tau)$ and $x_r(\tau)$ is quite analogous
to that of $\phi_d(\tau)$ and $\phi_r(\tau)$ of the previous section: $x_d
(\tau)$ is the solution to the equation of motion that stems from the action
(4.4):
$$
\dot x \ = \ x \, , \eqno \eq
$$
with the initial condition $x(0) = x_i$, and $x_r(\tau)$ is the solution to the
reverse (in time) equation
$$
\dot x \ = \ - \ x \, , \eqno \eq
$$
with the initial condition $x(0) = x_f$.

The distribution (4.7) was first obtained in Ref. \h\ by solving the stochastic
Langevin equation which corresponds to the Fokker-Planck equation (1.1). In
[\h] this expression was also
analysed. Here we shall perform similar brief analysis of the distributions
(4.7) and (4.8) to demonstrate excellent agreement
of the saddle-point approximation of the previous chapter with these exact
distributions. Using (4.1) and (4.2) we can express the distributions
(4.7) and (4.8) in terms of the scalar field variable $\phi$ as follows
$$\eqalign{
Z(\phi_f, \phi_i, \tau) \ & = \ {\cal N} \exp \left( - \ {3 M_{\rm P}^4
\left(\phi_f - \phi_d(\tau) \right)^2 \left( \phi_f + \phi_d(\tau)\right)^2
\over 8 \left(V_i - V_d(\tau)\right) \phi_f^4 } \right) \cr \cr
\ & = \ {\cal N} \exp \left( - \ {3 M_{\rm P}^4
\left(\phi_i - \phi_r(\tau) \right)^2 \left( \phi_i + \phi_r(\tau)\right)^2
\over 8 \left(V_r(\tau) - V_f \right) \phi_i^4 } \right) \, , \cr} \eqno \eq
$$
where we made notations $V_i = V(\phi_i)$, $V_f = V(\phi_f)$, $V_d(\tau) =
V(\phi_d(\tau))$, $V_r(\tau) = V(\phi_r(\tau))$.

It is straightforward to see that as long as the conditions (which are
essentially (1.10))
$$
{V_i - V_d(\tau) \over M_{\rm P}^4} \ \ll \ 1 \, , \eqno \eq
$$
and
$$
{V_r(\tau) - V_f \over M_{\rm P}^4} \ \ll \ 1 \, , \eqno \eq
$$
are valid, the distributions given by (4.11) are to a great precision
Gaussian with peak and variance in accord with the saddle-point approximation
of the previous chapter.
Thus, validity of our approach under the condition (1.10) is justified also by
comparison with an exact analytic solution.

{\bf \chapter{Path functional statistics}}

    The aim of this chapter is to  develop  methods  of  investigation
the  statistics of   path  functionals.  That  is,  we shall  learn
to calculate  mean   values $<F>$  of path functionals $F[\phi(\tau)]$.
In this paper we shall deal with the conditional statistics only,
the condition being that  the   scalar  field   passes
from $\phi_i$ to $\phi_f$ in a time $\tau_\ast$. If it is
known, statistics of any other type in principle can be easily determined.

Like  in  the  previous  chapters  we  restrict  ourselves  to  the
saddle-point approximation. In Chapter 2 we have seen that this approximation
is good enough for the scalar field probability distributions. For it to
be also good for the functional statistics, the functionals are
to vary sufficiently slowly (not exponentially) in the space of paths
$\phi(\tau)$, the condition which we shall assume valid.

Consider a set of path functionals $\{F_i\}$, $i = 1, 2, \ldots$ Then knowing
their  statistics is  equivalent  to
knowing the generating function $W$ of the set of real variables $\{\mu_i\}$
which is defined as follows
$$
\exp \left(- W(\mu) \right) \ = \ {\cal N} \int \exp \left( - I[\phi(\tau)] -
\sum_k \mu_k F_k [\phi (\tau)] \right) [d \phi] \, , \eqno \eq
$$
where by $\mu$ we shortly denoted the whole set $\{\mu_i\}$. Functional
integral in the last expression must be evaluated under an appropriate
condition; as it is clear from what has been emphasized above,  in  our
case this is the condition of the scalar field  passing  from $\phi_i$ to
$\phi_f$ in a time $\tau_\ast$. Normalization constant ${\cal N}$ in (5.1)
is supposed to be chosen in such a way that $W(0) = 0$. We shall use the
following properties of the function $W(\mu)$:
$$
{\partial W(0) \over \partial \mu_i} \ = \ <F_i> \, , \eqno \eq
$$
$$
{\partial^2 W(0) \over \partial \mu_i \partial \mu_j } \ = \ - \ <\delta F_i
\delta F_j > \, , \eqno \eq
$$
where $\delta F_i = F_i\, - <F_i>$.

In the saddle-point approximation the expression for $W(\mu)$ is given by
$$
W(\mu) \ \approx \ I(\mu) - I(0) + \sum_k \mu_k F_k (\mu) \, , \eqno \eq
$$
where $I(\mu)$ and $F_k (\mu)$ are the corresponding functionals' values on
the  extremal  path of the extended ``action''
$$
I_{\rm ext}\, [\phi(\tau)] \ \equiv \ I[\phi(\tau)] \ + \ \sum_k \mu_k F_k
[\phi(\tau)] \, . \eqno \eq
$$
All the expressions below will refer to the saddle-point  approximation.

{}From  the  above  definitions  it  is  clear  that
$$
<F_i>_{\mu \neq 0} \ = \ {\partial W(\mu) \over \partial \mu_i} \ = \
F_i(\mu) \, , \eqno \eq
$$
$$
<\delta F_i \delta F_j> \ = \ - \ {\partial F_i(0) \over \partial \mu_j} \
= \ - \ {\partial F_j(0) \over \partial \mu_i} \, . \eqno \eq
$$
Rewriting the last expression in somewhat other terms we obtain
$$
<\delta F_i \delta F_j> \ = \ - \ \int \limits_0^{\tau_\ast} d \tau\,
{\delta F_i
[\phi_c] \over \delta \phi (\tau)} {\partial \phi_c (\tau, 0) \over
\partial \mu_j} \, , \eqno \eq
$$
where $\phi_c (\tau, \mu)$ is the extremal path of the  extended action (5.5).
All
that  remains  to  do  now is to calculate the value $\partial \phi_c (\tau, 0)
/ \partial \mu_j$. Let us denote this value by $\beta_j (\tau)$:
$$
\beta_j (\tau) \ \equiv \ {\partial \phi_c (\tau, 0) \over \partial \mu_j}
\, . \eqno \eq
$$

Differential equation  for $\beta_j (\tau)$  can  be  obtained  from  the
equation of motion for $\phi(\tau, \mu)$. It can be put in the following form
$$
{d \over d \tau} \left( {\beta_j \over \dot \phi_c } \right) \ = \
{B(\phi_c) \over \dot \phi_c^2} \left( \int \limits_0^\tau d \sigma\,
{\delta F_j[\phi_c]
\over \delta \phi (\sigma)} \dot \phi_c (\sigma) + C_j \right)
\, , \eqno \eq
$$
where $C_j  =$ const. In this expression $\mu$ everywhere should  be  put  to
zero. The solution to the last equation is
$$
\beta_j (\tau) \ = \ \dot \phi_c (\tau) \int \limits_0^\tau d \sigma \,
{B(\phi_c(\sigma)) \over \dot \phi_c^2(\sigma)} \left( \int \limits_0^\sigma
d \rho\, {\delta F_j[\phi_c] \over \delta \phi(\rho)} \dot \phi_c(\rho) +
C_j \right) \, . \eqno \eq
$$
The constants $C_j$ are determined from the condition
$$
\beta_j (\tau_\ast) \ = \ 0 \,  ,   \eqno \eq
$$
which stems from our boundary condition $\phi(\tau_\ast, \mu) \equiv \phi_f$.

Now inserting the expression for $\beta_j (\tau)$ given by (5.11) into (5.8)
and
making use of the condition (5.12) we obtain our final result in the form
$$\eqalign{
<\delta F_i \delta F_j> \ = \ & \int \limits_0^{\tau_\ast} d \tau\,
{B(\phi_c(\tau)) \over \dot \phi_c^2(\tau)} f_i(\tau) F_j(\tau)\, - \cr
& - \,{\int \limits_0^{\tau_\ast} d \tau\, {B(\phi_c(\tau)) \over \dot
\phi_c^2(\tau)} f_i(\tau) \cdot \int \limits_0^{\tau_\ast} d \tau\,
{B(\phi_c(\tau)) \over \dot \phi_c^2(\tau)} f_j(\tau) \over
\int \limits_0^{\tau_\ast} d \tau\, {B(\phi_c(\tau)) \over \dot
\phi_c^2(\tau)}}\, , \cr} \eqno \eq
$$
where
$$
f_k (\tau) \ = \ \int \limits_0^\tau d \rho\, {\delta F_k[\phi_c] \over \delta
\phi(\rho)} \dot \phi_c(\rho) \, . \eqno \eq
$$

Let us consider a simple example. Let $F[\phi(\tau)] = \phi(\xi)$.
In this  case the formula (5.13)  yields
$$
<\delta \phi(\xi) \delta \phi (\eta)> \ = \ {\dot \phi_c(\xi) \dot \phi_c(\eta)
\int \limits_0^\eta d \rho\, {B(\phi_c(\rho)) \over \dot
\phi_c^2(\rho)} \cdot \int \limits_\xi^{\tau_\ast}
d \sigma {B(\phi_c(\sigma)) \over
\dot \phi_c^2(\sigma)} \over \int \limits_0^{\tau_\ast} d \tau {B(\phi_c(\tau))
\over \dot \phi_c^2(\tau)}} \, , \eqno \eq
$$
for $\eta \leq \xi$. It is easy to show that at $\eta = \xi$, and if
$\tau_\ast$ is the time in which the extremal trajectory defined by (3.1)
proceeds from $\phi_i$ to $\phi_f$, this expression
coinsides with the variance given by the equation (3.18). This,
of course, is of no surprise since in this case we calculate one and the same
value. If the time $\tau_\ast$ deviates strongly from the time in
which the extremal trajectory (3.1) passes from $\phi_i$ to $\phi_f$ then the
expression (3.18) for the variance ceases to be valid, and one has to use
more exact expression (5.15).

{\bf \chapter{Discussion}}

In this concluding chapter we shall briefly discuss possible  applications
of our
approach. To our mind the biggest advantage of the functional approach
is  in  the
possibility of dealing    with    the    statistics    of     the
inflationary-history-dependent  values.
Of course, this will be  of  essential  interest  only  for  {\it observable}
(directly or indirectly) values of such type. It is a general opinion
that in the context of the chaotic inflation no present observable value is
sensitive to the inflationary history in the very remote past [\l]. The reason
for this belief is that the spatial scales which correspond to the observable
structure are much
smaller than the Planck scale at most part of the inflationary history. They
become larger than the Planck scale only at the (relative) end of inflation
and it is
assumed that after this moment there is no ``memory'' on the spatial scales of
interest about the past evolution of the universe. However, this suggestion
is not so evident and might turn out to be not quite correct.
Thus in our work [\cs] we followed another assumption, namely, that
the quantum state of the universe at the end of inflation might contain
information
about the whole past evolution of the universe on the inflationary stage. We
assumed that the quantum state of the universe can be calulated using
the standard quantum field theory methods even on spatial scales much below the
Planck scale. Our result was that under these assumptions
stochastic inflation
can  lead  to  anisotropic  spectrum  of  the  primordial density fluctuations.
 The  degree of this anisotropy is a functional of  the  type discussed in the
present paper. In our work [\cs] the functional method developed in the present
paper was partially used. Although the value discovered in [\cs] is the only
known example of interest which depends substantially on the past inflationary
history, other possibilities may well appear in the future. It is for such
inflationary-history dependent functionals that the methods developed in this
paper will be of use.

\bigskip

{\bf Acknowledgments}

Most of the results of this work have been obtained in 1989 at Lebedev Physical
Institute in Moscow. I am very grateful
to G.V.Chibisov for many valuable discussions and to A.D.Linde for
encouragement to publish these results. Possibility of solving exactly the
model
with $\lambda \phi^4$ potential was communicated to me by V.F.Mukhanov,
to whom I am also very grateful. This work was also supported in part by
the National Research Council under the COBASE program.
\bigskip

\appendix

In this appendix we demonstrate how a path-integral solution to the diffusion
equation
(1.1) with $\alpha = 1/2$ (Stratonovich form) can be derived also from the
underlying Langevin equation using functional methods.

Consider  the  following  Langevin  equation
$$
\dot x \  = \ - \ A (x)  +  C (x) \eta \, , \eqno \eq
$$
with random  gaussian  function $\eta(t)$ representing white noise
$$
<\eta(t) \eta(u)> \ = \ \delta (t - u) \,  .  \eqno \eq
$$
The formal probability distribution in the space of functions $\eta (t)$
is  given by \Ref\f{R.F.Fox,  {\it Phys. Rev.} {\bf A33} (1985) 467.}
$$
\exp \left( - {1 \over 2} \int \eta^2(t)\, dt \right) \, . \eqno \eq
$$

Given a functional $F[x(t)]$ we can calculate its mean value:
$$
<F[x(t)]> \ = \ {\cal N} \int [d \eta]\, F[x_\eta(t)] \exp \left( - {1 \over 2}
\int \eta^2(t)\, dt \right) \, , \eqno \eq
$$
where $x_\eta (t)$ is the solution to (A.1), and the formal path  integration
measure in (A.4) is
$$
[d \eta] \ = \ \prod_t d \eta(t) \, . \eqno \eq
$$

If the initial  condition is  $x(0) = x_0$  and $F[x(t)] = \delta (x - x(t))$,
we  obtain  the ``Green's function''
$$
Z(x, x_0, t) \ = \ <\delta (x - x(t))> \, . \eqno \eq
$$

 It can  easily be shown  [\f]  that this function obeys  the  following
Fokker-Planck  equation
$$
{\partial Z(x, x_0, t) \over \partial t} \ = \ {\partial \over \partial x}
\left( A(x)Z(x, x_0, t) +
{1 \over 2} C(x) {\partial \over \partial x} C(x) Z(x, x_0, t) \right)
\, . \eqno \eq
$$
This equation is just the equation (1.1) with $B = C^2$, $\alpha = 1/2$,
written in terms of the variable $x$ instead of $\phi$.

It is natural to proceed in (A.4) from  integration  over  $\eta(t)$  to
integration over $x(t)$. We will have then
$$
<F[x(t)]> \ = \ {\cal N} \int [dx]\, F[x(t)] \exp \left(- I[x(t)]
\right) \, . \eqno \eq
$$
The ``action'' $I[x(t)]$ in this expression is given by (compare with (2.1))
$$
I[x(t)] \ = \ {1 \over 2} \int dt\, {\left(\dot x + A(x) \right)^2 \over
C^2(x)} \, , \eqno \eq
$$
and the formal measure for path integration is found to be
$$
[dx] \ = \ \prod_t {dx(t) \over C(x(t))} \, . \eqno \eq
$$

We see that (A.8), (A.9) and (A.10) coincide respectively with (1.8), (1.6)
and (1.7).

\bigskip

\par
\refout
\end